\pdfoutput=1
\documentclass[12pt]{iopart}
\jl{33}
\usepackage{graphicx}
\usepackage{color}
\usepackage{url}

\newcommand{\eqref}[1]{{(\ref{#1})}}
\newcommand{\tfrac}[2]{{\textstyle\frac{#1}{#2}}}

\newcommand{\D}{\mathrm{d}}
\newcommand{\bfP}{{\mathbf{P}}}
\newcommand{\bfp}{{\mathbf{p}}}
\newcommand{\bfe}{{\mathbf{e}}}
\newcommand{\bfr}{{\mathbf{r}}}
\newcommand{\bfv}{{\mathbf{v}}}
\newcommand{\pdp}{{\bfp\cdot\delta\bfp}}
\newcommand{\dpdp}{{\delta\bfp\cdot\delta\bfp}}

\begin{document}

\title[Polar active liquids: a universal classification~\ldots]
      {Polar active liquids: a universal classification rooted in nonconservation of momentum}
\author{Khanh-Dang Nguyen Thu Lam, Michael Schindler and Olivier Dauchot}
\address{UMR Gulliver 7083 CNRS, ESPCI ParisTech, PSL Research University, 10 rue Vauquelin, 75005 Paris, France}
\date{21 October 2015}
\vspace{5mm}
\begin{indented}
\item Received 8 May 2015
\item Accepted for publication 14 September 2015
\item Published 21 October 2015
\end{indented}
\vspace{5mm}
\begin{indented}
\item Journal of Statistical Mechanics: Theory and Experiment (2015) P10017
\item Online at \url{http://stacks.iop.org/JSTAT/2015/P10017}
\item \url{doi:10.1088/1742-5468/2015/10/P10017}
\end{indented}

\begin{abstract}
We study the spatially homogeneous phases of polar active particles in the low
density limit, and specifically the transition from the isotropic phase to
collective polar motion. We show that the fundamental quantity of interest for
the stability of the isotropic phase is the forward component of the momentum
change induced by binary scattering events. Building on the Boltzmann formalism,
we introduce an ansatz for the one-particle distribution and derive a
closed-form evolution equation for the order parameter. This approach yields a
very intuitive and physically meaningful criterion for the destabilization of
the isotropic phase, where the ansatz is exact. The criterion also predicts
whether the transition is continuous or discontinuous, as illustrated in three
different classes of models. The theoretical predictions are in excellent
agreement with numerical results.
\end{abstract}

\maketitle

\section{Introduction}

Polar active liquids are composed of aligning self-propelled particles which
convert energy into directed motion. They generically exhibit large scale
collective motion~\cite{Ramaswamy:2010bf,Marchetti:2013bp}. Simulations of
Vicsek-like models of constant-speed point particles, aligning with their
neighbors in the presence of noise, have revealed the existence of a transition
between an isotropic phase and a true long-range order polar phase with giant
density fluctuations~\cite{vicsek1995novel,czirok1997spontaneously,
Gregoire:2004ic,Chate:2008is,ginelli2010relevance,Vicsek:2012ty}. For metric
interactions---with a density-dependent rate of collisions---the homogenous
polar state is unstable close to the transition; propagative structures develop
and the transition becomes discontinuous. An intense theoretical effort towards
the understanding of the long range behavior of these systems has lead to the
picture of a basic universality class, at least for the simplest situation in
which the surrounding fluid can be neglected (dry flocking) and the sole
interaction is some local
alignment~\cite{Toner:2005bj,toner1995long,toner1998flocks,Bertin2006,Bertin2009,
Ihle:2011ds,Chou:2012wq,Peshkov:2012uu,Peshkov2014,Ihle:2014tz,Caussin:2014te}.

However, Vicsek-like models contain some level of coarse-graining of the
dynamics and as such are not just ``simple liquids''~\cite{Hansen:105413}. For a
given system of particles, be it
experimental~\cite{Kudrolli:2008cd,Deseigne:2010gc,Palacci:2010hk,Theurkauff:2012ui,
Deseigne2012,Bricard:2013jq,Palacci:2013eu,Kumar:2014wr} or
numerical~\cite{peruani2006nonequilibrium,Grossman:2008is,Henkes:2011ed,Fily:2012tf,
Redner:2012vr}, it is thus crucial to check whether it indeed belongs to the
above universality class. This question has been addressed in a very limited
number of experimental situations only. In the case of rolling
colloids~\cite{Bricard:2013jq}, for which the hydrodynamics equations can be
derived explicitly, the interactions mediated by the surrounding fluid actually
dominate the alignment mechanism, but also screen the splay instability
responsible for giant density fluctuations in the polar phase. In the case of
walking grains~\cite{Deseigne:2010gc,Deseigne2012}, the alignment mechanism
results from complex re-collisional dynamics, and large-scale simulations reveal
some qualitative differences with the above canonical
scenario~\cite{Weber:2013bj}.

In some sense, both the complexity of the dynamics close to the transition, and
the technicality of the derivation of the hydrodynamic equations have hindered a
more basic question: is there a simple way to predict the existence and the
order of a transition to collective motion for a given microscopic dynamics? In
this letter, we tackle this question, restricting ourselves to the study of the
homogeneous phases of two-dimensional polar active liquids in the low density
limit. In such systems, the total momentum is changed by binary scattering and
self-diffusion events. We start from the Boltzmann equation formalism, assuming
that the molecular chaos hypothesis holds. With no further assumptions, we first
derive an evolution equation for the total momentum. However, this evolution
equation depends on the unknown angular distribution of the particle velocities.
We then propose an ansatz for this distribution, and obtain a closed-form
equation for the order parameter. Applying this equation in the isotropic phase,
where the ansatz is exact, we introduce a physically meaningful \emph{effective
alignment}, which is simply the average over all binary scattering events of the
nonconserved part of the momentum, projected onto the momentum before scattering.
The transition to collective motion occurs when this effective alignment is
larger than the disaligning effect of self-diffusion. A similar criterion also
predicts whether the transition is continuous or discontinuous. Finally, we test
and illustrate our approach on (i)~a mean-field Vicsek-like model (ii)~a
continuous-time model of hard disks obeying Vicsek aligning rules when
colliding, actually an implementation of the BDG
model~\cite{Bertin2006,Bertin2009}, and (iii)~a model of self-propelled
inelastic hard disks. In all cases, not only is the transition point very well
predicted, but the ansatz also works surprisingly well, even far into the polar
phase.

In the light of the important role played by inhomogeneous solutions, focusing
on the transition between homogeneous phases may look a bit academic. However,
revisiting the transition towards collective motion in terms of phases and phase
separations has recently proven to be an insightful
approach~\cite{Solon:2013vr,Solon:2014tu,Romenskyy2014}. Furthermore, following
the experimental discovery of topological interactions---with a
density-independent rate of collision---in bird
flocks~\cite{Ballerini2007}, it was shown that such systems remain
homogeneous across the
transition~\cite{ginelli2010relevance,Chou:2012wq,Peshkov:2012uu,
Degond:2013wt,Albi:2013ve}. Also, experimental systems of interest may have
small enough sizes such that homogeneous phases are stable. Finally, we shall
see that following this route leads us towards a very intuitive understanding of
the conditions which particle interaction must satisfy to induce a transition
towards collective motion.

\section{Theoretical framework}

Particle velocities at equilibrium obey the Maxwell--Boltzmann distribution;
self-propelled particles do not. After some transient, a self-propelled particle
reaches its intrinsic steady velocity $v_0$, set by the competition between
propelling and dissipation mechanisms~\cite{Deseigne2012,Bricard:2013jq,Hanke2013}.
In the low-density limit, this transient lasts much less time than the mean free
flight time, and one can safely assume that particles have a constant speed
$v_0$. For spatially homogeneous states, the one-particle distribution thus
reduces to the density probability $f(\theta,t)$ of having a particle with
velocity $v_0\hat\bfe(\theta)$ at time~$t$, where $\hat\bfe(\theta)$ is the unit
vector of polar angle~$\theta$. This distribution evolves according to
self-diffusion events and binary scattering events. It is crucial to clearly
specify what is meant by a binary scattering event, or rather, scattering
sequence: it begins when two particles start interacting and ends when they
recover their speed $v_0$. One should realize that (i) it can be rather complex,
involving, for instance, successive recollisions, as in systems of hard
disks~\cite{Deseigne2012}, (ii) even if the collision itself conserves momentum
as, for instance, in inelastic collisions, the intrinsic self-propulsion
dynamics enforces the particles to recover their steady velocity $v_0$ after the
collision, keeping the memory of the collision geometry, and thereby destroys
momentum conservation.
Hence, in general, the scattering of two self-propelled particles does not
conserve the average momentum of the system $\bfP(t) = \int\D\theta
f(\theta,t)\, \hat\bfe(\theta)$. Taking $\psi(t) = |\bfP(t)|$ as the order
parameter of the transition towards polar collective motion, it is thus natural
to analyze the change of momentum \emph{at the level of binary scattering}. As
we shall see, this allows us to understand collective macroscopic states,
starting from a microscopic description.

\subsection{Kinetic equations}

Within our approximations, the evolution equation for $f(\theta,t)$ is given by
the Boltzmann-like equation~\cite{Bertin2009}:
\begin{equation}
    \frac{\partial f}{\partial t}(\theta,t)
    =
    I_\mathrm{scat}[f,f] + I_\mathrm{diff}[f] ,
\end{equation}
where the binary scattering contribution is given by the scattering integral
$I_\mathrm{scat}[f,f]$ and where $I_\mathrm{diff}[f]$ describes the
self-diffusion process. The self-diffusion process, usually absent from the
Boltzmann equation, describes random kicks that the particle can either receive
from the medium on which it self-propels (e.g. as vibrated polar disks) or
generate by itself (e.g. run and tumble motion of bacteria). An integration of
this equation over $\theta$, using $\int\!\D\theta\, \hat\bfe(\theta)$, leads to
a kinetic equation for $\bfP(t)$. Here, we find it more instructive to obtain
such a kinetic equation by using an equivalent but more elementary derivation.

A scattering event, as pictured schematically on figure~\ref{fig:pdp}(left), is
specified by the incoming angles $\theta_1$ and $\theta_2$ of the two particles
or, equivalently, by the incoming half-angle \hbox{$\bar\theta =
\mathrm{Arg}(e^{i\theta_1}{+}e^{i\theta_2})$} and the incoming angular
separation \hbox{$\Delta = \theta_1{-}\theta_2$}. Additional scattering
parameters, such as the impact parameter, or some collisional noise, may be
needed and are collectively noted as~$\zeta$. A scattering event changes the
momentum sum of the two particles involved by an amount~$\delta\bfp$, which
depends \emph{a priori} on all scattering parameters $\bar\theta$, $\Delta$ and
$\zeta$. The average momentum of all~$N$ particles in the system changes in this
event from $\bfP$~into $\bfP'$, concluding that $N(\bfP'-\bfP)=\delta\bfp$. In
the same way, a self-diffusion event changes the momentum of a particle at
$\theta_1$ by an amount $N(\bfP'-\bfP)=\delta\bfp_\mathrm{diff}(\theta_1,\eta) =
\mathbf{R}_\eta \bfp - \bfp$, where $\mathbf{R}_\eta \bfp$ is the rotation of
$\bfp=\hat\bfe(\theta_1)$ by an angle $\eta$. The self-diffusion process is
characterized by the probability density~$P_\eta(\eta)$ for a particle with
angle $\theta_1$ to jump to angle $\theta_1+\eta$. Assuming molecular chaos and
averaging these two balance equations over the statistics of scattering and
self-diffusion events taking place in a small time interval, one obtains the
evolution equation by taking the continuous time limit:
\begin{figure}[t]
\begin{center}
    \includegraphics[width=5cm]{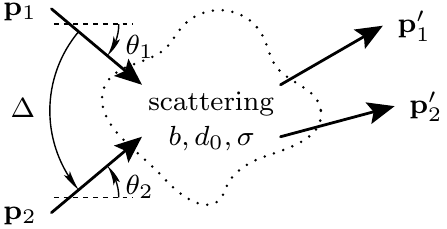}
    \hspace{1em}
    \includegraphics[width=4.5cm]{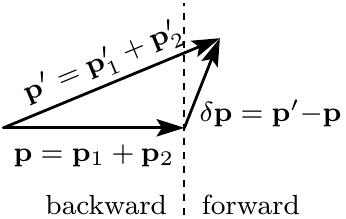}
\end{center}
\caption{
    \emph{Left:} scattering of two particles.
    \emph{Right:}
    criterion for stability of the isotropic phase.
    The momentum of two interacting particles is changed from $\bfp$
    to~$\bfp'$. If $\bfp'$~is more likely to be found in the forward
    semi-plane, the isotropic state is unstable.
}%
\label{fig:pdp}%
\end{figure}%
\begin{equation}
\label{eq:kin}
\label{eq:kin1}
    \frac{\D\bfP}{\D t}
    =
    \lambda\:
    \Phi^{\mathrm{scat}}_f\Bigl[\delta\bfp(\bar\theta,\Delta,\zeta)\Bigr]
     + \;
     \lambda_\mathrm{diff}\:
     \Phi^{\mathrm{diff}}_f\Bigl[\delta\bfp_\mathrm{diff}(\theta_1,\eta)\Bigr] ,
\end{equation}
where
\begin{eqnarray}
\label{eq:kin2}
\Phi^{\mathrm{scat}}_f\Bigl[\dots\Bigr]
&=
    \int_0^{2\pi}  \D\bar\theta
    \int_{-\pi}^{\pi} \D\Delta
    \int \D\zeta\,
    K(\Delta,\zeta) \, f(\theta_1,t)f(\theta_2,t)\,
    (\dots)
,\\
\label{eq:kin3}
\Phi^{\mathrm{diff}}_f\Bigl[\dots\Bigr]
&=
    \int_0^{2\pi} \D\theta_1
    \int \D\eta \, P_\eta(\eta) \, f(\theta_1,t)\,
    (\dots) .
\end{eqnarray}%
In the right hand side of equation~\eqref{eq:kin1}, the second term comes from
the self-diffusion process, which happens at a characteristic rate
$\lambda_{\mathrm{diff}}$. The first term comes from the binary scattering
process. In its integrand, a scattering event with scattering parameters
$\theta_1$, $\theta_2$ and $\zeta$ is assumed to happen at a rate proportional
to both $f(\theta_1,t)$ and $f(\theta_2,t)$; this comes from the molecular chaos
hypothesis. The proportionality factor is $\lambda K(\Delta,\zeta)$, the
scattering rate of such an event. Note that it does not depend on $\bar\theta$
as a result of global rotational invariance. As a convention, we have chosen to
normalize $K$ such that $\frac{1}{2\pi}\int_{-\pi}^{\pi} \D\Delta \int\D\zeta
K(\Delta,\zeta) = 1$. The prefactor $\lambda$ thus gives the characteristic
scale of the scattering rate. In what follows, we shall consider two cases.

(i) Nonmetric models: in systems of flying flocks, interaction between birds is
not defined in terms of a metric distance, but rather in terms of a topological
one~\cite{Ballerini2007}: the birds interact with a fixed number of their
nearest neighbors, at a rate $\lambda$, regardless of the distance between the
two interacting particles and their angular
separation~\cite{ginelli2010relevance}. Another motivation for studying this
kind of model concerns the physics of mean-field-like models, where interactions
are defined by a random quenched network~\cite{aldana2007phase}. For this class
of models, $\lambda$ is a free parameter and $K(\Delta,\zeta)=K(\zeta)$ does not
depend on $\Delta$.

(ii) Metric models: if one considers interacting disks with diameter $d_0$ at a
density number $\rho$, a scattering event is entirely described by $\theta_1$,
$\theta_2$ and the impact parameter $b$ (thus, $\int\D\zeta\equiv
\int_{-d_0}^{d_0}\D b$). By using the construction of the Boltzmann
cylinder~\cite{Kardar}, one finds for the scattering rate $\lambda K(\Delta,b) =
\rho v_0 |\sin\frac\Delta2|$. Importantly, it is proportional to the density and
does not depend on the impact parameter. The Boltzmann cylinder expresses the
fact that tangential scattering (small $|\Delta|$) occurs at a lower rate than
frontal scattering (large $|\Delta|$). Indeed, in tangential scattering,
particles are more parallel and, having the same speed, have a smaller relative
velocity, hence a lower scattering rate. On the other hand, particles have a
higher relative velocity in frontal scattering, hence a higher scattering rate.

Equation~\eqref{eq:kin} gives the evolution of the vectorial order parameter
$\bfP$. Now, in order to get the evolution of $\psi=|\bfP|$, we go to polar
coordinates $\bfP = \psi\,\hat\bfe(\theta_P)$ and project
equation~\eqref{eq:kin1} onto the radial direction~$\hat\bfe(\theta_P)$.
When the scattering and self-diffusion processes obey the mirror symmetry (no
chirality), $\bfP$ keeps its angular direction so that one can set
$\theta_P(t)=0$. As for the binary scattering term, we find for the projection
$\Phi^{\mathrm{scat}}_f \bigl[\delta\bfp\bigr] \cdot \hat\bfe(\theta_P) =
\Phi^{\mathrm{scat}}_f\bigl[(\hat\bfp \cdot \delta\bfp)\cos\bar\theta\bigr]$.
For the self-diffusion term, we can compute the integral explicitly and obtain
$\lambda_\mathrm{diff}\Phi^{\mathrm{diff}}_f\bigl[\delta\bfp_\mathrm{diff}\bigr]
= -D\psi$, where the self-diffusion constant is given by
\begin{equation}
D=\lambda_\mathrm{diff}\left(1-\int\D\eta\,P_\eta(\eta)\cos\eta\right) \ge 0 .
\end{equation}
It is instructive to look at an angular noise with zero expectaction and
variance $\sigma_0^2$. Using $P_\eta(\eta) =
\exp(-\eta^2/2\sigma_0^2)/\sqrt{2\pi\sigma_0^2}$, one finds
$D=\lambda_\mathrm{diff}(1-e^{-\sigma_0^2/2})$. In particular, when the angular
noise is weak, $\sigma_0\ll1$, one has $D\propto \sigma_0^2$.
Altogether, the radial component of equation~\eqref{eq:kin1} reads:
\begin{equation}
    \label{eq:dpsidt}
    \frac{\D\psi}{\D t}
    =
    \lambda \Phi^\mathrm{scat}_f\!\Big[(\hat\bfp \cdot\delta\bfp)\cos\bar\theta \Big] - D\psi .
\end{equation}
This evolution equation is derived from equation~\eqref{eq:kin} with the only
additional assumption being that the system is not chiral. We keep this
assumption in what follows.

\subsection{The von Mises distribution ansatz}

\begin{figure}[t]
\begin{center}
    \includegraphics[]{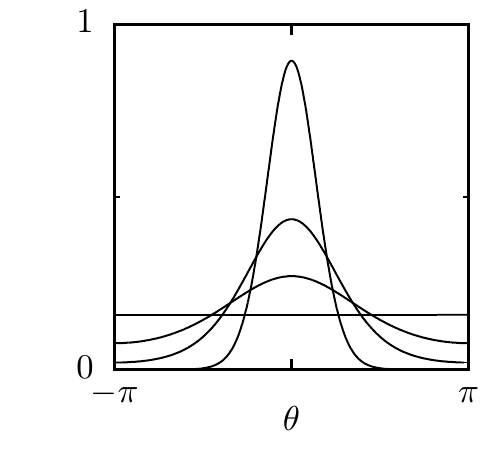}
    \hspace{1em}
    \includegraphics[]{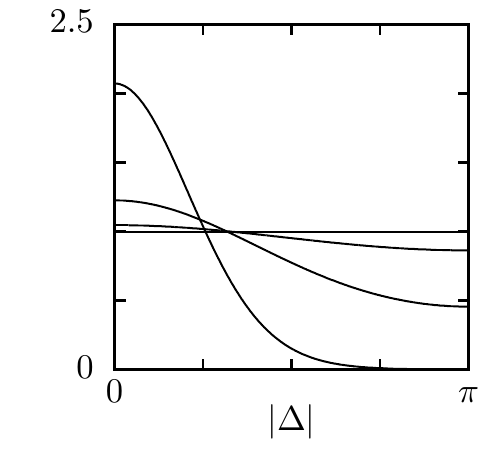}
\end{center}
\caption{
    \emph{Left:} the von Mises distribution $f_\psi(\theta)$, for
    $\psi=0$, $0.3$, $0.6$ and $0.9$.
    \emph{Right:}
    the kernel $g(\psi,\Delta)/\psi$ for the same values of $\psi$.
}%
\label{fig:vm}%
\end{figure}%

The above kinetic equations remain of limited pratical interest as long as the
angular distribution~$f$ is unknown. Here, we propose an ansatz of the form
$f(\theta,t)=f_{\psi(t)}(\theta)$, which we constrain to be exact in the
isotropic phase. We choose $f_\psi$~to be the so-called von Mises
distribution~\cite{Watson1982}, the distribution of random angles, uniform up to
the constraint $\bigl|\!\int\!\D\theta
f_\psi(\theta)\hat\bfe(\theta)\bigr|=\psi$. This distribution maximizes the
entropy functional $H[f]=-\int f\log{f}$ under the aforementioned constraint and
is, in this sense, the simplest ansatz one can think of and was actually used to
study Vicsek-like models~\cite{Degond2013,Chepizhko2014}. It is parameterized by
the order parameter $\psi$ in the following way:
\begin{equation}
    f_\psi(\theta) = \frac{e^{\kappa(\psi) \cos\theta}}{2\pi I_0(\kappa(\psi))},
    \quad
    \mathrm{with}
    \quad
    \frac{I_1(\kappa)}{I_0(\kappa)} = \psi,
\label{eq:vonmises}
\end{equation}
where $I_n(x)$ is the modified Bessel function of the first kind, of order $n$.
Plots of this distribution for different values of $\psi$ are available in
figure~\ref{fig:vm}(left). In the limits $\psi\to0$ ($\kappa\to0$) and $\psi\to1$
($\kappa\to\infty$), one recovers respectively the isotropic distribution
$f(\theta)=1/2\pi$ and a normal distribution of variance $1/\kappa$. For all
values of $\psi$ (equivalently of $\kappa$), this distribution has a single
maximum at $\theta=0$ and a single minimum at $\theta=\pm\pi$. It is more peaked
as $\psi$ or $\kappa$ is higher. The symmetry $\theta\leftrightarrow-\theta$
expresses the nonchirality of the system. After injecting this ansatz into
equation~\eqref{eq:dpsidt}, the integration over $\bar\theta$ can be performed
analytically. Because the ansatz is parameterized by $\psi$, one obtains a
closed-form equation for the evolution of $\psi$:
\begin{equation}
\label{eq:dpsidt_Ansatz}
    \frac{1}{\lambda}\frac{\D\psi}{\D t} =  F(\psi) - \frac{D}{\lambda}\psi,
\end{equation}
The binary scattering term $F(\psi)$ is a nonlinear function of $\psi$ and a
functional of the scattering function $\pdp(\Delta,\zeta)$:
\begin{equation}
\label{eq:F_psi}
    F(\psi)
    = \int_{-\pi}^{\pi}\! \frac{\D\Delta}{2\pi}
       \int\! \D\zeta\,
       K(\Delta,\zeta) \,
       g(\psi,\Delta) \,
       \pdp(\Delta,\zeta)
,
\end{equation}
where
\begin{equation}
    g(\psi,\Delta)
    =
\frac{\kappa(\psi)}{I_0\bigl(\kappa(\psi)\bigr)^2}
\frac{I_1\bigl(2\kappa(\psi)\cos\frac{\Delta}{2}\bigr)}{2\kappa(\psi)
    \cos\frac{\Delta}2}
.
\end{equation}
The kernel $g(\psi,\Delta)$ is plotted in figure~\ref{fig:vm}(right). Its
behaviour can be interpreted in the following way. As the system is more polar
(higher $\psi$), it is more likely to find pairs of aligned particles than
anti-aligned particles: scattering at low values of $|\Delta|$ is favored as
compared to scattering at $|\Delta|\simeq \pi$. Around the isotropic state
$\psi\ll1$, one has $g(\psi,\Delta)\simeq \psi$, which, as expected, does not
depend on $\Delta$. The accuracy of the ansatz, and hence the accuracy of $g$,
are tested numerically below.

\subsection{Instability of the isotropic state: a proper definition of the alignment}

To account for the destabilization of the isotropic state, one must consider the
spontaneous fluctuations of $\psi\ll1$. At linear order in $\psi$, the von Mises
distribution reads
\begin{equation}
    f_\psi(\theta) = \frac1{2\pi} \big( 1 + 2\psi \cos\theta \big) .
\end{equation}
Note that this distribution verifies the self-consistency condition
$|\!\int\D\theta f_\psi(\theta)\hat\bfe(\theta)|=\psi$. This is the distribution
one would obtain, assuming that the orientations of the particles are not
correlated, the most reasonable assumption for the isotropic phase. It is in
this sense that the ansatz can be said to be exact for the description of the
isotropic state. At linear order in $\psi$, equation~\eqref{eq:dpsidt_Ansatz} reads
\begin{equation}
\label{eq:isotropic}
   \frac{1}{\lambda} \frac{\D\psi}{\D t} = \mu \psi,
\end{equation}
where
\begin{equation}
\label{eq:mu}
   \mu = \langle \pdp \rangle_0 - \frac{D}{\lambda}
\end{equation}
with
\begin{equation}
\label{eq:average0}
    \langle\dots\rangle_0
    = \int_{-\pi}^{\pi}\! \frac{\D\Delta}{2\pi}
       \! \int\! \D\zeta\,
       K(\Delta,\zeta) \,
       (\dots) .
\end{equation}
The above set of equations is our central result. It is exact within the
approximations of the Boltzmann equation for nonchiral systems. In particular,
it does not rely on the choice of the ansatz for the angular distribution. As we
shall see now, it provides an intuitive understanding of when polar collective
motion develops in systems of polar active particles and allows us to define
properly the alignment of scattering events. The isotropic state is stable when
$\mu<0$ and unstable when $\mu>0$, while solving for $\mu=0$ gives the
transition. The sign of $\mu$ is set by two terms in equation~\eqref{eq:mu}. The
first one is the average of the change of momentum in the forward direction
($\pdp$) over the space of scattering parameters. As we will see, it can be of
either sign, depending on the details of the interactions. Note that the average
defined in equation~\eqref{eq:average0} is conveniently normalized such that
$\langle1\rangle_0=1$. In the second term of equation~\eqref{eq:mu}, the
self-diffusion noise $D\ge0$ acts on the scale of the free flight time
$1/\lambda$ and has the effect of destroying polar order. For metric models, the
interaction rate scales as $\lambda \propto \rho$. In this case, solving for
$\mu=0$ leads to the somewhat trivial linear dependence of the critical
diffusion coefficient with density $D_c \propto \rho$ (or $\sigma_{0c}\propto
\sqrt\rho$), as commonly reported in the literature.

In equations~\eqref{eq:dpsidt_Ansatz} and \eqref{eq:isotropic}, all the
model-specific microscopic details of the interaction between particles appear
only through the forward momentum change $\pdp$. The latter is positive when
$\delta\bfp$ points forward, i.e. in the same ``direction'' as $\bfp$,
see figure~\ref{fig:pdp}(right). It is often said that a scattering event
``aligns'' particles when it decreases the angular separation between the
velocities, that is when $\left|\bfp'\right| \!>\!\left|\bfp\right|$. However,
it is easy to see from figure~\ref{fig:pdp}, that this microscopic alignment
property is a necessary condition for having $\pdp>0$, although not a sufficient
one, since a large enough angular deviation of momentum can always bring $\bfp'$
in the backward semi-plane. We learn here that $\pdp$~is the proper quantity to
evaluate the microscopic alignment taking place in a scattering event. It allows
us to write the linear coefficient in equation~\eqref{eq:mu} in a more compact and
meaningful form than previously obtained general expressions, see equation (35)
in~\cite{Peshkov2014}. We have checked that the integrand in that equation is
actually equal to $\pdp$.

The mechanism of the instability of the isotropic state is clear: if there is
some fluctuation of polar order $\psi\neq0$, the momentum of two interacting
particles is statistically more likely to be found along the direction of this
fluctuation. Then, if $\langle\pdp\rangle_0>D/\lambda$, momentum is created on
average along this same direction by binary scattering, building polar order
faster than the self-diffusion noise is able to destroy it.

\subsection{Nature of the transition}

To predict whether the transition is continuous or discontinuous, we go beyond
linear order, expanding equation~\eqref{eq:dpsidt_Ansatz} up to order $\psi^3$:
\begin{equation}
\label{eq:order3}
	\frac{1}{\lambda} \frac{\D\psi}{\D t} = \mu \psi - \xi \psi^3,
\end{equation}
where
\begin{equation}
\label{eq:xi}
    \xi = \langle (\tfrac12 - \cos\Delta) \, \pdp\rangle_0.
\end{equation}
If $\xi>0$ at the transition, the transition is continuous and the polar state
$\psi \simeq \sqrt{\mu/\xi}$ emerges continuously as a new stable stationary
state. If $\xi<0$, the transition is discontinuous and one must expand
equation~\eqref{eq:dpsidt_Ansatz} to higher orders in~$\psi$ to compute the new
stable stationary state. The expression in equation~\eqref{eq:xi} depends on the
ansatz for the angular distribution. However, the sign is what matters for the
prediction of the nature of the transition. One can show that a continuous
transition is indeed predicted as such by equation\eqref{eq:xi},
see~\ref{annex:aboutxi}.

In equation~\eqref{eq:xi}, the factor $\frac12-\cos\Delta$ gives a negative
contribution for tangential scattering (low angles $|\Delta|$), and a positive
contribution for frontal scattering (large angles $|\Delta|$). In models where
tangential scattering dis-aligns while frontal scattering aligns, the transition
is prone to be continuous. We will see below that models with interactions given
by the Vicsek collision rule fall in this class of models. On the other hand, in
models where tangential scattering mostly aligns and where frontal scattering
mostly dis-aligns, coefficient $\xi$ is more likely to be negative. There is
thus the propensity for this kind of model to display a discontinuous
transition rather than a continuous one. As we will see below, in a model of
self-propelled hard disks with inelastic collisions, this qualitative argument
gives a correct prediction.

\subsection{Fluctuations of the order parameter}

We can obtain information on the fluctuations of the order parameter by
computing the value of $\psi^2=\bfP^2$ in the stationary state. One way to do
it is to start again from the momentum balance equation $N(\bfP'-\bfP) =
\delta\bfp$. Taking the square, we obtain the balance equation
\begin{equation}
    \label{eq:balanceP2}
    N(\bfP'^2 - \bfP^2) = 2\bfP\cdot\delta\bfp + \frac1N\dpdp,
\end{equation}
from which one can obtain a kinetic equation for $\bfP^2$, using the same
derivation as presented above. Finally, making use of the von Mises
distribution ansatz, one can obtain a closed-form evolution equation. The
derivation can be found in \ref{annex:fluct}. As it is not particularly
instructive, we present here only the main result. In the isotropic state, the
variance of $\bfP$ is given by
\begin{equation}
\label{eq:varP_iso}
\textrm{Var}[\bfP] =
    \langle \bfP^2 \rangle = \frac1N \frac{
        \tfrac12 \langle\dpdp\rangle_0 + D/\lambda
    }{|\mu|
    }
    .
\end{equation}
It scales as $1/N$ as expected. In the numerator, the fluctuations arise both
from the fluctuations of $\delta\bfp$ in binary scattering events and from the
fluctuations of the self-diffusion process. The denominator $|\mu|$ is the
``restoring force'', which vanishes at the transition. The absolute value comes
from $\mu$ being negative in the isotropic phase. In the isotropic phase, $\bfP$
follows a Gaussian distribution and it is easy to show that
$\langle\psi\rangle^2 = \tfrac\pi4 \langle\bfP^2\rangle$, from which one obtains
the variance of the scalar order parameter
\begin{equation}
\langle\psi^2\rangle - \langle\psi\rangle^2 = (1-\tfrac\pi4)\langle \bfP^2 \rangle .
\end{equation}
This prediction is in full agreement with numerical measurements in the three
models studied below.

\section{Application to models}

We now come to the illustration of these mechanisms in the cases of three
different models. We also test numerically the accuracy of the von Mises ansatz.
We focus the discussion on the binary scattering properties, illustrating the
link between the alignment function $\int_\zeta \pdp$ of the models and the
corresponding collective behaviour. We thus study the models without any
self-diffusion noise by setting $D=0$. As we described quantitatively by
equation~\eqref{eq:mu}, the $D>0$ case shifts the transition by stabilizing the
isotropic phase.

\subsection{Mean-field binary Vicsek model}
\label{sec:bdgmf}

We first consider a nonmetric model where interactions are binary, with a
change of momentum that follows the collision rule of the Vicsek model. At every
time step, two randomly chosen particles among $N\gg1$ collide, following the
binary Vicsek collision rule: from precollision velocity angles $\theta_1$ and
$\theta_2$, the half-angle
$\bar\theta=\mathrm{Arg}(e^{i\theta_1}+e^{i\theta_2})$ is computed and randomly
rotated to $\bar\theta+\eta_1$ and $\bar\theta+\eta_2$. The collisions' noises
$\eta_1$ and $\eta_2$ are two independent noises following a Gaussian
distribution of variance $\sigma^2$,
$P(\eta)=e^{-\eta^2/2\sigma^2}/\sqrt{2\pi\sigma^2}$. The two new angles are then
assigned to the unit velocity vectors of the particles. The collision noise
$\sigma$ is used as the control parameter. As for the Vicsek model, it has the
effect of blurring out the alignment to the half-angle $\bar\theta$ and we expect an
isotropic phase at large $\sigma$ and a polar phase at small $\sigma$. An
important difference with the Vicsek model, apart from the absence of space, is
that interactions are only binary, whereas particles in the Vicsek model can
interact through multiple interactions.%
\begin{figure}[t]
    \begin{center}
    \includegraphics[]{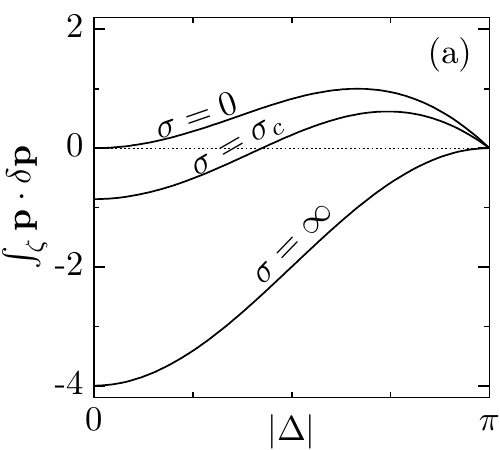}
    \hspace{.5cm}
    \includegraphics[]{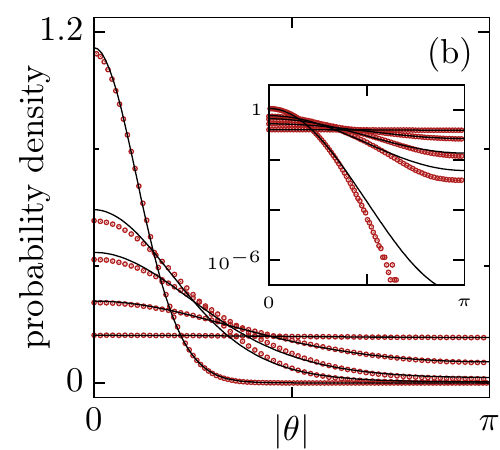}
    \hfill
    \includegraphics[]{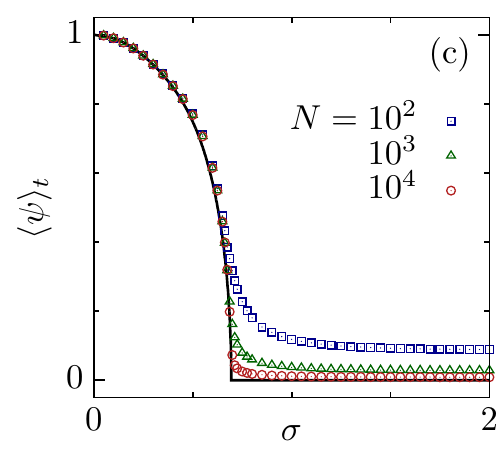}
    \hspace{.5cm}
    \includegraphics[]{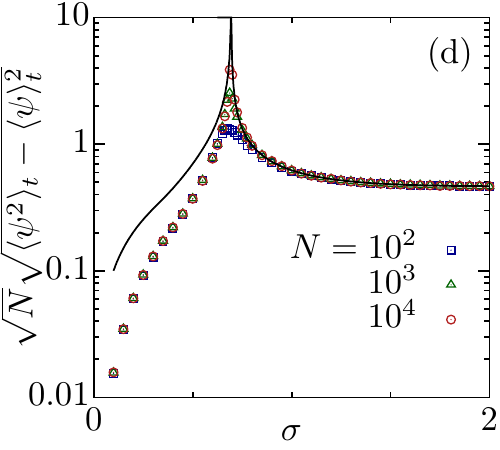}
    \end{center}
\caption{
    Mean-field binary Vicsek model.
    (a)
    The alignment function $\int_\zeta\!\pdp$,
    for different values of the control parameter $\sigma$.
    (b)
    Angular distributions measured at $N=10^4$ (symbols) and
    the corresponding ansatz distributions (lines). From top to
    bottom: $\sigma=0.25,0.5,0.6,0.675,1$ (respectively,
    $\psi\simeq 0.94,0.77,0.61,0.32,0.01$).
    Inset: the same but with vertical log-scale.
    (c), (d)
    Average in the steady state of the order parameter and its rescaled standard
    deviation. Symbols are numerical solution of the Boltzmann equation. Full
    black lines are theoretical predictions using the von Mises distribution as
    an ansatz for the angular distribution.
}
\label{fig:bdgmf}
\end{figure}

The model is termed as a mean-field one, as a particle can interact with any
other one, with no correlation of any kind. By construction, the molecular chaos
hypothesis holds exactly for this model, as the master equation that describes
the dynamics is exactly the Boltzmann equation. Thus, the discrepancy between
the theoretical predictions and the numerical data comes only from finite-size
effects, which are negligable here as we will see, and from the inaccuracy of
the ansatz. This kind of mean-field models thus provides a way to test for the
accuracy of the ansatz in a controlled way. Also, we will compare the results
for the present model to a similar one, sharing the same collision rule, but
where binary interactions depend on a metric distance. As we will see, the
physics of the transition is qualitatively the same for both models.

Let us first look at the theoretical predictions. It is easy to see that
$\pdp=|\bfp|(\cos\eta_1+\cos\eta_2-|\bfp|)$, where $|\bfp|=2\cos\tfrac\Delta2$.
The integration over the collision noises is performed using $\int\D\zeta \equiv
\int\D\eta_1 \D\eta_2 P(\eta_1) P(\eta_2)$. We obtain the alignment function
\begin{equation}    \label{eq:pdp_bdg}
    \int \D\zeta \, \pdp
    = 2\cos\tfrac{\Delta}{2} \,
    \Bigl( 2e^{-\sigma^2/2} - 2\cos\tfrac{\Delta}{2} \Bigr).
\end{equation}
This function of the incoming angular separation $\Delta$, represented in
figure~\ref{fig:bdgmf}(a), summarizes the microscopic dynamics averaged over the
``internal'' degrees of freedom of the scattering (here the collision noise):
for $\sigma=0$ it is always positive, all collisions align on average; for
$\sigma=\infty$ it is always negative, there is no alignment on average. At
intermediate $\sigma$, collision with a large, respectively small, incoming
angle separation $\Delta$ align, respectively dis-align. Computing the
coefficient $\mu$ now simply consists of averaging this function against the
kinetic kernel $K$. Here, there is no spatial dependence of any kind, and $K$ is
just a constant. Using equations~\eqref{eq:mu} and \eqref{eq:xi}, we integrate
equation~\eqref{eq:pdp_bdg} over $\int\D\Delta$, obtaining $\mu=\tfrac8\pi
e^{-\sigma^2/2} - 2$ and $\xi=\tfrac4{3\pi} e^{-\sigma^2/2}$. Solving for
$\mu=0$, the transition occurs at $\sigma_c=\sqrt{2\log(4/\pi)}\simeq 0.695$
and, because $\xi(\sigma_c)>0$, the transition is continuous. To extend the
predictions to the polar phase, we set $\D\psi/\D t=0$ in
equation~\eqref{eq:dpsidt_Ansatz} and solved it numerically, to obtain the order
parameter. These predictions are presented in figure~\ref{fig:bdgmf} in full
black lines.

We compare them to numerical results obtained using the following Monte Carlo
method~\cite{Bird1970}. Starting from $N$ random angles $\theta_i(t)$, a pair of
distinct particles $(i,j)$ is chosen randomly, uniformly. The collision rule is
then applied, obtaining the new angles $\theta_i(t+1)$ and $\theta_j(t+1)$. All
other particles keep their angle. The procedure is repeated until the stationary
state is reached. We then start to measure averages over time of quantities of
interest. In our simulations, these averages typically involved $10^6$
collisions, which gives us good enough statistics. Finite-size effects in the
simulations are under control, as shown by the scaling in $N$ in
figure~\ref{fig:bdgmf}(d). Quite remarkably, the measured angles distributions
compare well with the ansatz in the whole range of $\psi$, see
figure~\ref{fig:bdgmf}(b). Time averages of the order parameter in the stationary
state also compare very well with the theoretical prediction in the whole range
of $\psi$, see figure~\ref{fig:bdgmf}(c). Concerning the fluctuations of the order
parameter, one must distinguish the isotropic phase from the polar one. In the
isotropic phase the predictions are excellent, see figure~\ref{fig:bdgmf}(d),
confirming that the correlations are negligible. In the polar phase, the von
Mises distribution ansatz is not supposed to be exact, which translates into a
qualitative agreement only. Finally, increasing the size of the system, the
divergence of the fluctuations at the transition is better and better captured.

\subsection{Continuous-time hard disks Vicsek model}

We next consider a metric model, with $N$ hard disks of diameter $d_0=1$ moving
in a periodic box of linear size $L$. The number density is $\rho=N/L^2$. In
this model, speeds are fixed to $v_0=1$. As we do not consider self-diffusion,
particles go in a straight line until a collision occurs. Two particles interact
when $|\bfr_1-\bfr_2|=d_0$, their velocities are changed following the binary
Vicsek collision rule, as already defined in the previous model (alignment to
the half-angle and a collision noise with variance $\sigma^2$). A way to ensure
that the interaction is always binary is to prevent particles from overlapping. This
is achieved by noticing that there is only one way to assign the two outcoming
velocities to the two particles, out of the two possibilities. We choose to
assign the velocities such that particle do not overlap, hence the terming of
\emph{hard} disks. Using this rule, the interaction between particles is binary
and the interaction is made instantaneous, which has allowed us to define a
continuous-time model. The current model, in the dilute regime where molecular
chaos hypothesis holds, is an actual implementation of the one studied
theoretically in~\cite{Bertin2006,Bertin2009}, but not simulated therein.
By comparison, in the original Vicsek model, the dynamics are discrete in time.
As a consequence, particles behave like disks that can overlap at any time with
one or many other particles. Note also that, in the Vicsek dynamics, a
scattering event, even if binary, can last many time steps.%
\begin{figure}[t]
    \begin{center}
    \includegraphics[]{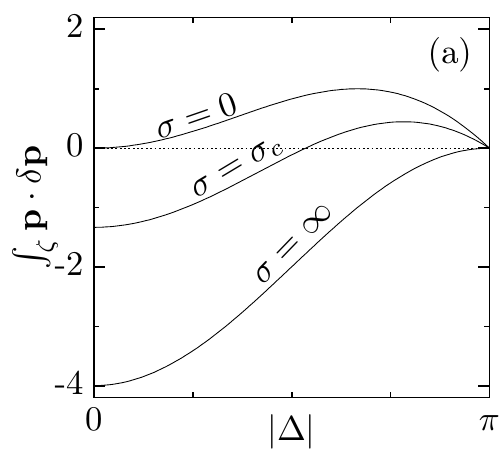}
    \hspace{.5cm}
    \includegraphics[]{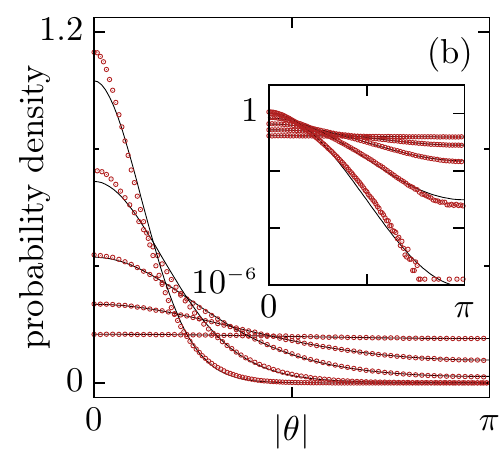}
    \hfill
    \includegraphics[]{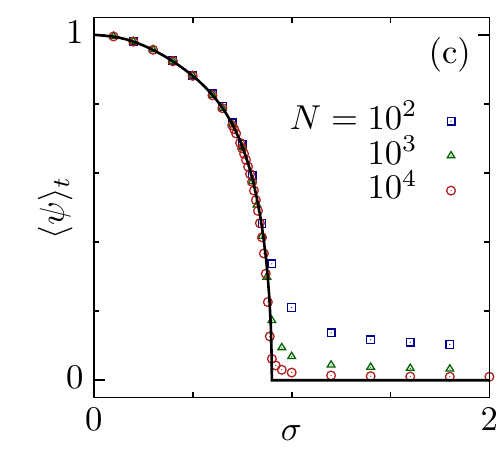}
    \hspace{.5cm}
    \includegraphics[]{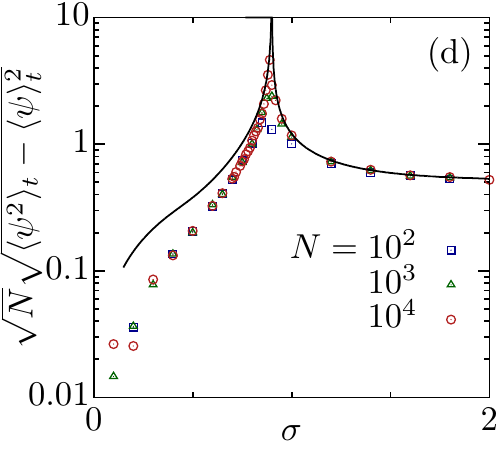}
    \end{center}
\caption{
Continuous-time hard disks Vicsek model at density $\rho=10^{-3}$.
    {(a):}
    The alignment function $\int_\zeta\!\pdp$,
    for different values of the control parameter $\sigma$.
    {(b):}
    Angular distributions measured at $N=10^4$ (symbols) and
    the corresponding ansatz distributions (lines). From top to
    bottom: $\sigma=0.4, 0.6, 0.8, 0.88, 1$ (respectively
    $\psi\simeq 0.93, 0.83, 0.59, 0.30, 0.02$).
    Inset: the same but with vertical log-scale.
    {(c), (d):}
    Average in the steady state of the order parameter and its rescaled
    standard deviation.
    Symbols are numerical data from molecular dynamics simulations.
    Full black lines are theoretical predictions.
}
\label{fig:bdg}
\end{figure}

\begin{figure}[t]
    \begin{center}
    \includegraphics[]{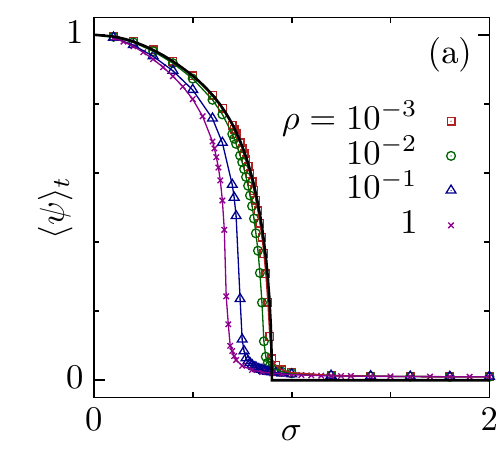}
    \hspace{.5cm}
    \includegraphics[]{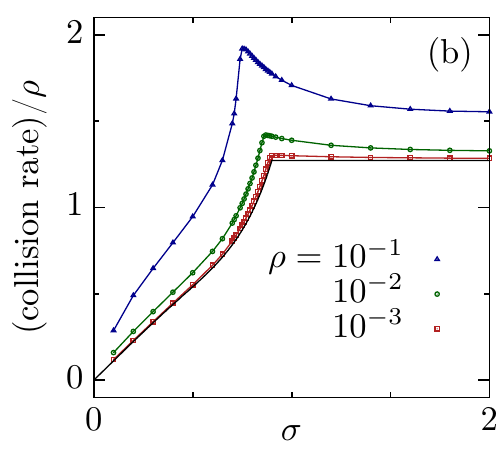}
    \end{center}
\caption{
    Continuous-time hard disks Vicsek model
    at densities $\rho=10^{-3}$, $10^{-2}$, $10^{-1}$ and $1$.
    (a) Polar order parameter.
    (b) Collision rate, rescaled.
    For clarity, the data at $\rho=1$ is not plotted. At this density, the
    rescaled collision rate behaves qualitatively the same as for lower
    densities, but with much higher values (reaching around $28$ for the maximal
    value at the transition). Symbols are numerical data, $N=10^4$.
    Black lines are theoretical predictions at vanishing density.
}
\label{fig:bdg:rho}
\end{figure}

Here again, the collision noise $\sigma$ is used as a control parameter. The
collision rule being the same as for the mean-field Vicsek model, the alignment
function $\int_\zeta\!\pdp$ is \emph{the same} as equation~\eqref{eq:pdp_bdg}. For
the theoretical description, the only difference stems from the kinetic kernel,
which reads here $K(\Delta)\propto\left|\sin(\Delta/2)\right|$, as given by the
construction of the Boltzmann cylinder. We also remind the reader that the scattering rate
is set by the density, $\lambda \propto \rho$. Again one can compute $\mu$,
following equation~\eqref{eq:mu}, obtaining $\mu=\frac8\pi e^{-\sigma^2/2} -
\frac{16}{3\pi}$, which cancels at $\sigma_c=\sqrt{2\log(3/2)}\simeq0.9005$. We
recover the same results as in~\cite{Bertin2006,Bertin2009}. We also find
that the transition is continuous, $\xi(\sigma_c)>0$. Remember that this
statement only concerns the transition between homogeneous states. It does not
rule out the discontinuous transition scenario reported for this system, which
involves the destabilization of the homogeneous polar phase with respect to
inhomogeneous solutions~\cite{Bertin2006,Bertin2009}. Finally, we can also solve
numerically equation~\eqref{eq:dpsidt_Ansatz} for the order parameter in the polar
phase, see the black lines in figures~\ref{fig:bdg} and
\ref{fig:bdg:rho}(a).

We obtained numerical data in the stationary state of molecular dynamics
simulations. In the absence of noise, we used an event-driven method, which allowed
us to probe more easily the low density regime.
Some minimal care has to be taken, as the spatial homogeneity of the stationary
states can be destroyed by hydrodynamic instabilities. Practically, these
instabilities are known to occur at quite large
wavelengths~\cite{Chate:2008is,Ihle:2011ds}, so that we used small-sized
systems. We checked explicitly that the simulations run in the homogeneous
regime. In particular, no travelling bands were observed in our simulations,
even for the largest system $N=10^4$. Here also, the theoretical predictions are
in very good agreement with the simulation data at low density $\rho=10^{-3}$,
see figure~\ref{fig:bdg}.

The current model is similar to the mean-field binary Vicsek model defined in
the previous section: the binary interaction obeys the same collision rule. Only
the scattering rates' dependance on $\Delta$ differs. Comparing
figures \ref{fig:bdgmf}~and \ref{fig:bdg}, both models share the same
qualitative behaviours. In particular, the transition is continuous in both
cases. One sees that at the level of homogeneous phases, the nature of the
transition is clearly governed by the collision rule rather than by the
metric/nonmetric aspect of the interaction.%

We also investigated finite density effects on the order parameter and the
collision rate, as shown in figure~\ref{fig:bdg:rho}. Although they are hardly
seen at density below $\rho=10^{-3}$, deviations become more and more noticeable
as density increases. For the order parameter (figure~\ref{fig:bdg:rho}(a)), an
increase in density \emph{stabilizes} the isotropic phase. This is in contrast
with the most commonly reported effect of stabilization of the polar phase by
density, in the presence of self-diffusion. In the present case, there is no
self-diffusion ($D=0$) and the transition shift comes from truly nontrivial
correlations. For the collision rate, a quantity most easily measured in
event-driven simulations, see figure~\ref{fig:bdg:rho}(b), a prediction can be
obtained by computing $\lambda\Phi_f^\mathrm{scat}[1]$, using the von Mises
distribution ansatz. The idea is simply to count $+1$ at each collision, instead
of $\delta\bfp$ in kinetic equations such as equation~\eqref{eq:kin1}. The
result is plotted as a black full line in figure~\ref{fig:bdg:rho}(b). In the
isotropic phase, the collision rate is simply the constant $\lambda =4\rho/\pi$.
In the polar phase, it decreases smoothly as $\psi$ is increased. This is again
a pure kinetic effect. When polar order is higher, particles are more parallel,
with smaller relative velocities, so it takes more time before a collision is
likely to occur. The collision rate vanishes for $\psi=1$, when all particles
are strictly parallel. From the numerical data, we observe first that the
overall collision rate is increased as density gets higher; second, that for a
given density the collision rate increases as the transition is approached from
either side, reaching a finite maximal value at the transition. While the first
feature is expected, as it also happens in equilibrium
systems~\cite{Hansen:105413}, the second one indicates a nontrivial dependance
of the collision rate with density in the transitional regime. These effects
cannot be understood on the basis of the Boltzmann formalism.

\subsection{Inelastic self-propelled hard disks}

Several works~\cite{Grossman:2008is,Lobaskin2013,Coburn2013} have shown that
pairwise dissipative interactions lead to global polarization in swarms of SPPs.
In the present model, particles are hard disks of diameter $d_0=1$ that collide
inelastically. The restitution coefficient $0\le e\le1$ of the inelastic
collisions is used as a control parameter for the transition. Between
collisions, the dynamics of particle $i$ is given by
\begin{eqnarray}
&\frac{\D\bfr_i}{\D t} &= \bfv_i ,\\
\label{eq:bar}
\tau &\frac{\D\bfv_i}{\D t} &= \mathrm{sign}(v_0-|\bfv_i|) \hat\bfv_i ,
\end{eqnarray}
where $\mathrm{sign}(x)$ is $-1$, $0$ or $1$, respectively, when $x$ is negative, zero
or positive. The rhs term of equation~\eqref{eq:bar} allows us to use event-driven
methods to perform molecular dynamics simulations. It mimics the more standard
exponential relaxation of the velocity $\bfv_i$ to $\hat\bfv_i=\bfv_i/|\bfv_i|$
on a timescale $\tau$. We also studied the case of an exponential relaxation,
though in less details, for which we observe that all the results presented
below are qualitatively the same. We choose $v_0=1$ and $\tau=1$.
\begin{figure}[t]
    \begin{center}
    \includegraphics[]{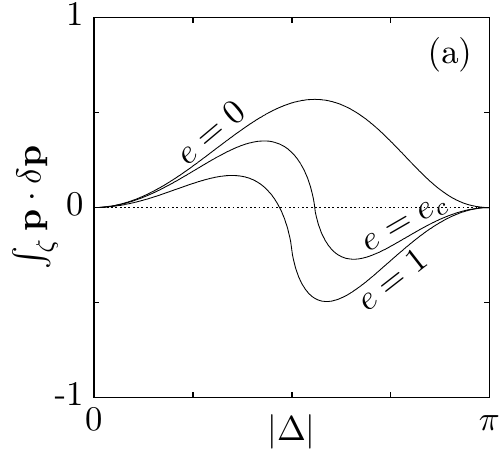}
    \hspace{.5cm}
    \includegraphics[]{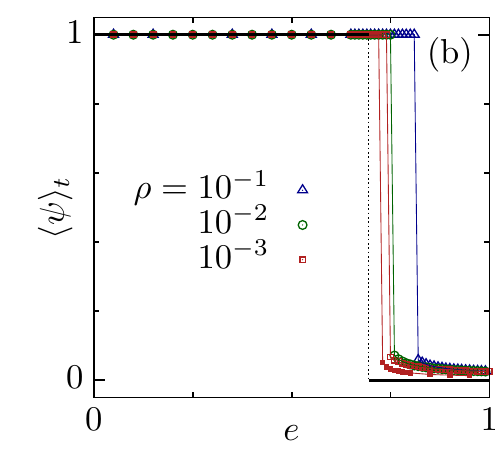}
    \end{center}
\caption{
    Self-propelled hard disks model with inelastic collisions, $\tau=1$.
    (a)
    The alignment function $\int_\zeta\!\pdp$,
    for different values of the control parameter $e$.
    (b) Polar order parameter.
    Symbols: numerical data, $N=1000$ (open symbols), $N=4000$ (full symbols).
    Black lines are theoretical predictions at vanishing density.
}
\label{fig:inelasticspp}
\end{figure}

For this model, the $\int_\zeta\!\pdp$ functions are computed numerically by
simulating many binary scattering events at some fixed incoming angular
separation $\Delta$, varying the impact parameter $b$ uniformly, see
figure~\ref{fig:inelasticspp}(a). Here, as already stated in the theoretical
framework section, the distinction between \emph{binary scatterings events} and
\emph{binary collisions} is particularly important. A binary scattering event
starts at the time of a first collision, when both particles have speed $v_0$,
with a momentum $\bfp$. After some time, the particles separate forever and the
dynamics restore the speed of both particles to $v_0$. Only when all these
conditions are eventually met does the binary scattering event end and we record
the momentum $\bfp'$. We insist that while the momentum is conserved by
inelastic \emph{collisions}, it is not by the scattering event; the reason being
that after the collision, velocities are being relaxed to $v_0$ and momentum is
changing, so that in general $\bfp'\neq\bfp$. Note also that a single
binary scattering event can comprise several inelastic collisions, depending on
the parameters of the scattering. From these data, we can compute
$\pdp(b,\Delta)$, then $\mu$ and $\xi$, using equations \eqref{eq:mu}
and~\eqref{eq:xi}. We find a transition at $e_c\simeq 0.70$. Because
$\xi(e_c)<0$, the transition is predicted to be discontinuous. The results are
in full agreement with direct molecular dynamics simulations with a random
isotropic state as initial conditions. As shown in
figure~\ref{fig:inelasticspp}(b), the transition is indeed highly discontinuous.

Around the transition $e=e_c$, as seen in figure~\ref{fig:inelasticspp}(a),
tangential collisions (low $|\Delta|$) align, while frontal collisions (high
$|\Delta|$) disalign. This is in total contrast with the binary Vicsek collision
rule, see figures \ref{fig:bdgmf}(a) and \ref{fig:bdg}(a). Note that tangential
collisions align for all values of $e$. As a consequence, the fully polar state
$\psi=1$ is stable for all values of $e$. Indeed, when $\psi\simeq1$, particles
are all quite parallel, so that binary scattering only occurs at low $|\Delta|$.
In this scattering regime, $\pdp>0$, so that polar order increases back to
$\psi=1$. There is thus a coexistence of stability between the $\psi=1$ and
$\psi=0$ states, hence a discontinuous transition. Note that one could define a
mean-field like version of this model, by considering the model in
section~\ref{sec:bdgmf}, but with a collision rule given by inelastic
collisions, figure~\ref{fig:inelasticspp}(a), instead of the Vicsek collision
rule, figure~\ref{fig:bdgmf}(a). The results would be qualitatively the same,
with a sharp discontinuous transition. Here again, the quantity $\pdp$ is more
important with respect to the nature of the transition than the kinetic kernel.
As a final remark, here, as opposed to the previous model, higher densities tend
to stabilize the polar phase, even in the absence of self-diffusion.

\section{Conclusion}

In summary, proposing an ansatz for the velocity angular distribution, we have
derived an equation for the evolution of the momentum of systems of polar active
particles with fixed speed. The weakly nonlinear analysis around the isotropic
state is given by equation~\eqref{eq:isotropic} and provides an intuitive way of
anticipating the transition to collective motion in systems of polar active
particles: the existence and the nature of the transition are essentially
governed by the way $\int_\zeta\!\pdp$ depends on the incoming angle. As an
important consequence, the forward component of momentum change, $\pdp$, is the
proper quantity to characterize the alignment of binary scattering. Also, we
tested the fully nonlinear equation on three different kinds of models, and
showed that the von Mises ansatz describes quite well the velocity angular
distribution, even for large polarization. This shows that knowing the value of
the order parameter, alone, already gives much qualitative information about the
kinetics. Of course, predictions of more subtle effects in the polar phase
should require better approximation schemes. These encouraging results naturally
call for the extension of our analysis to models in which the particle speeds
are free to fluctuate. Our work may also be adapted to describe the transition
towards nematic states or three dimensional systems.

\section*{Acknowledgements}
The authors would like to thank E Bertin for enlightening discussions.

\appendix

\section{On the sign of the cubic term}
\label{annex:aboutxi}

As the derivation of equation~\eqref{eq:xi} involves the use of an ansatz, it is
not ``exact'' and one should worry about the sign of $\xi$ being wrong. Let us
now compare the expression of equation~\eqref{eq:xi} with the one obtained
in~\cite{Peshkov2014}, where the starting point is also the Boltzmann
equation, but where the hydrodynamics equations are derived using a different
closure scheme. We first briefly describe how these equations are obtained.
Starting from the Fourier series $f(\theta,t) = \frac1{2\pi}\sum_k f_k(t) e^{-i
k\theta}$, the Boltzmann equation can be written in Fourier space. The result is
an infinite number of coupled equations: the time evolution of the $k$th mode is
given as a function of the other modes. Next, the following scaling hypothesis
is assumed close to the transition~\cite{Peshkov2014}: $|f_k| \sim
\epsilon^{|k|}$ and $\partial_t \sim \epsilon$, for some small parameter
$\epsilon$. Neglecting all contributions of order $\epsilon^4$ and those of
higher order, the Boltzmann equation reduces to~\cite{Peshkov2014}:
\begin{eqnarray}
    \partial_t f_0 &= 0 , \\
    \label{eq:foo1}
    \partial_t f_1 &= \mu_1 f_1 - \xi_1 f_1^* f_2 , \\
    \label{eq:foo2}
    \partial_t f_2 &= \mu_2 f_2 + \gamma f_1^2 ,
\end{eqnarray}
The first equation states that a homogeneous density field stays homogeneous.
The second one is the analog of our evolution equation for the vectorial order
parameter $\bfP$. The third equation describes the evolution of the nematic
order parameter. We have discarded the equation for $\partial_t f_3$ because
$f_1$ and $f_2$ do not depend on $f_3$. As long as the scaling hypothesis holds
uniformly, these equations can be considered as ``exact''. Note that they apply
only in the case $\xi_1>0$, since $\xi_1<0$ would require the inclusion of
higher order terms.

In what follows, we suppose that the isotropic state is linearly stable with
respect to the nematic phase, hence $\mu_2<0$, and consider the slightly polar
state, $\mu_1>0$ with $\mu_1\simeq 0$. As $|f_1|\neq0$, we are free to choose
the reference direction by setting $f_1^*=f_1 > 0$. In the stationnary state,
equations~\eqref{eq:foo1} and~\eqref{eq:foo2} are equated to zero, so as to obtain
\begin{equation}
    |f_1|^2 = \frac{\mu_1}{\xi_1}
    \frac{|\mu_2|}\gamma
,
\qquad
\qquad
    \label{eq:foof2}
    f_2 =
        \frac\gamma{|\mu_2|} f_1^2
.
\end{equation}
We see from the first equation that one must have $\gamma>0$ and from the second
that $f_2>0$. The expression of the stationary $f_2$ in equation~\eqref{eq:foof2}
can be used in equation~\eqref{eq:foo1}, which then reads
\begin{equation}
\label{eq:foofoo}
\partial_t f_1 =
    \mu_1 f_1 - \xi_1 \frac{\gamma}{|\mu_2|} f_1^* f_1^2 .
\end{equation}
We see that the ``exact'' cubic coefficient is given by $\xi_1 \gamma/|\mu_2| >
0$. Note that it has the same sign as $\xi_1>0$.

We now come back to our results, obtained from the von Mises distribution
ansatz. The expansion in powers of $f_1$ of the ansatz in
equation~\eqref{eq:vonmises} reads
\begin{equation}
f_\psi(\theta) =
\tfrac1{2\pi}
+ \tfrac1\pi (
    {\psi} \cos\theta
    + {\psi^2} \cos{2\theta}
    + \dots
) .
\end{equation}
One important difference here is that the second Fourier mode is enslaved to the
first one, such that $f_2=f_1^2$, instead of having equation~\eqref{eq:foof2}.
Thus, one finds $\partial_t f_1 = \mu_1 f_1 - \xi_1 f_1^3$, instead of
equation~\eqref{eq:foofoo}. By identifying this equation with
equation~\eqref{eq:order3}, it is shown that
(i) $\mu_1=\mu$, as given by equation~\eqref{eq:mu},
(ii) $\xi_1=\xi$, as given by equation~\eqref{eq:xi},
(iii) the cubic term in equation~\eqref{eq:order3} has the correct sign.
Thus, if the transition predicted by equation~\eqref{eq:foofoo} is continuous,
equation~\eqref{eq:order3} also predicts a continuous transition: both
approaches are consistent. Interestingly, this mainly comes from $f_2$ and $f_1$
sharing the same sign for a continuous transition, which means that both the
polar mode and the nematic modes are in phase, a property also possessed by the
von Mises distribution.

When $\xi_1<0$, one has to consider higher order ``exact'' equations.
Unfortunately, even the order~7 equations are not well
behaved~\cite{Peshkov2014}.

\section{Fluctuations of the order parameter}
\label{annex:fluct}

Here, we derive an expression for the variance of the order parameter. We start
from equation~\eqref{eq:balanceP2}, the balance equation for the momentum:
\begin{equation}
    N(\bfP'^2 - \bfP^2) = 2\bfP\cdot\delta\bfp + \frac1N\dpdp.
\end{equation}
Assuming that the system is nonchiral, we can follow the procedure already used
for deriving equation~\eqref{eq:kin1}. We find
\begin{equation}
\frac1\lambda \frac{\D\bfP^2}{\D t}
=
{2\psi}\,
    \Phi_f^\mathrm{scat} \Big[\hat\bfp\cdot\delta\bfp \cos\bar\theta\Big]
+ \frac2N
    \Phi_f^\mathrm{scat} \Big[\dpdp\Big]
- 2 \frac{D}\lambda \Big(\psi^2 - \frac1N \Big)
,
\end{equation}
with $\Phi_f^\mathrm{scat}$ defined in Eq.~\eqref{eq:kin2}. Note that this
equality stands at the level of ensemble average. Using the von Mises
distribution ansatz, the integration over $\bar\theta$ can be performed
explicitly and this expression becomes
\begin{equation}
\label{eq:dpsi2dt_Ansatz}
\frac1\lambda \frac{\D\bfP^2}{\D t}
=
{2\psi} F(\psi)
+ \frac2N G(\psi) - 2 \frac{D}\lambda \Big(\psi^2 - \frac1N \Big) ,
\end{equation}
where $F(\psi)$ is already given by Eq.~\eqref{eq:F_psi} and where
\begin{equation}
\label{eq:G_psi}
    G(\psi)
    = \int_{-\pi}^{\pi}\! \frac{\D\Delta}{2\pi}
       \int\! \D\zeta\,
       K(\Delta,\zeta) \,
   \frac{I_0\Big(2 \kappa(\psi) \cos\tfrac\Delta2\Big)}{I_0^2(\kappa(\psi))}
\,
       \tfrac12\dpdp(\Delta,\zeta)
.
\end{equation}
Now, consider the ensemble averaged stationary state $\bfP_*$, and the
trajectory of the system around this average: $\bfP(t)=\bfP_* + \delta\bfP(t)$,
with $\delta\bfP(t)$ assumed to be of order $1/\sqrt{N}$. The order 0 of
equation~\eqref{eq:dpsi2dt_Ansatz} gives the condition for the stationary state,
$F(\psi_*) - (D/\lambda)\psi_* = 0$, while orders $1/\sqrt{N}$ and $1/N$ are
respectively
\begin{eqnarray}
     \frac1{\lambda}
         \frac{\D \bfP_*\cdot\delta\bfP}{\D t}
    &=
    \left( F'(\psi_*) - \frac{D}{\lambda} \right)\bfP_*\cdot\delta\bfP
    ,
\\
    \frac1{2\lambda}
        \frac{\D (\delta\bfP)^2}{\D t}
    &=
    \left( F'(\psi_*) - \frac{D}\lambda \right) (\delta\bfP)^2
    +
    \frac1N \left( G(\psi_*) + \frac{D}{\lambda} \right)
    ,
\end{eqnarray}
where $F'(\psi)\equiv{\D F}/{\D\psi}$. The first equation is the stability
condition of the stationary state, thus requiring that
$F'(\psi_*)<\frac{D}\lambda$. Equating the second equation to zero, we get the
variance of the order parameter in the stationary state:
\begin{equation}
\textrm{Var}[\bfP]
    =
(\delta\bfP)^2
    = - \frac1{N} \frac{
        G(\psi) + D/\lambda
   }{
           F'(\psi_*) - D/\lambda
    }
    .
\end{equation}
The minus sign comes from the denominator being negative. Note that this
expression is not expected to be quantitatively accurate in the polar phase. In
the isotropic state, this expression becomes equation~\eqref{eq:varP_iso}.

\section*{References}
\bibliographystyle{iopart-num}

\begin{thebibliography}{10}
\expandafter\ifx\csname url\endcsname\relax
  \def\url#1{{\tt #1}}\fi
\expandafter\ifx\csname urlprefix\endcsname\relax\def\urlprefix{URL }\fi
\providecommand{\eprint}[2][]{\url{#2}}

\bibitem{Ramaswamy:2010bf}
Ramaswamy S 2010 {\em Annu Rev Conden Ma P\/} {\bf 1} 323--345

\bibitem{Marchetti:2013bp}
Marchetti M~C, Joanny J~F, Ramaswamy S, Liverpool T~B, Prost J, Rao M and Simha
  R~A 2013 {\em Rev. Mod. Phys.\/} {\bf 85} 1143--1189

\bibitem{vicsek1995novel}
Vicsek T, Czir{\'o}k A, Ben-Jacob E, Cohen I and Shochet O 1995 {\em Phys. Rev.
  Lett.\/} {\bf 75} 1226--1229

\bibitem{czirok1997spontaneously}
Czir{\'o}k A, Stanley H~E and Vicsek T 1997 {\em Journal of Physics A:
  Mathematical and General\/} {\bf 30} 1375

\bibitem{Gregoire:2004ic}
Gr{\'e}goire G and Chat{\'e} H 2004 {\em Phys. Rev. Lett.\/} {\bf 92} --

\bibitem{Chate:2008is}
Chat{\'e} H, Ginelli F, Gr{\'e}goire G and Raynaud F 2008 {\em Phys. Rev. E\/}
  {\bf 77} --

\bibitem{ginelli2010relevance}
Ginelli F and Chat{\'e} H 2010 {\em Phys. Rev. Lett.\/} {\bf 105} 168103

\bibitem{Vicsek:2012ty}
Vicsek T and Zafeiris A 2012 {\em Physics Reports\/} {\bf 517} 71--140

\bibitem{Toner:2005bj}
Toner J, Tu Y and Ramaswamy S 2005 {\em Annals of Physics\/} {\bf 318} 170--244

\bibitem{toner1995long}
Toner J and Tu Y 1995 {\em Phys. Rev. Lett.\/} {\bf 75} 4326--4329

\bibitem{toner1998flocks}
Toner J and Tu Y 1998 {\em Phys. Rev. E\/} {\bf 58} 4828--4858

\bibitem{Bertin2006}
Bertin E, Droz M and Gr{\'e}goire G 2006 {\em Phys. Rev. E\/} {\bf 74} 22101

\bibitem{Bertin2009}
{Bertin} E, {Droz} M and {Gr{\'e}goire} G 2009 {\em Journal of Physics A
  Mathematical General\/} {\bf 42} 445001

\bibitem{Ihle:2011ds}
Ihle T 2011 {\em Phys. Rev. E\/} {\bf 83} 030901

\bibitem{Chou:2012wq}
Chou Y~L, Wolfe R and Ihle T 2012 {\em Phys. Rev. E\/} {\bf 86} 021120

\bibitem{Peshkov:2012uu}
Peshkov A, Ngo S, Bertin E, Chat{\'e} H and Ginelli F 2012 {\em Phys. Rev.
  Lett.\/} {\bf 109} 098101

\bibitem{Peshkov2014}
Peshkov A, Bertin E, Ginelli F and Chat\'e H 2014 {\em The European Physical
  Journal Special Topics\/} {\bf 223} 1315--1344 ISSN 1951-6355
  \urlprefix\url{http://dx.doi.org/10.1140/epjst/e2014-02193-y}

\bibitem{Ihle:2014tz}
Ihle T 2014 {\em Eur. Phys. J. Special Topics\/} {\bf 223} 1293--1314

\bibitem{Caussin:2014te}
Caussin J~B, Solon A, Peshkov A, Chat{\'e} H, Dauxois T, Tailleur J, Vitelli V
  and Bartolo D 2014 {\em Phys. Rev. Lett.\/} {\bf 112} 148102

\bibitem{Hansen:105413}
Hansen J~P and McDonald I~R 1976 {\em {Theory of simple liquids}\/} (London:
  Academic Press)

\bibitem{Kudrolli:2008cd}
Kudrolli A, Lumay G, Volfson D and Tsimring L~S 2008 {\em Phys. Rev. Lett.\/}
  {\bf 100}

\bibitem{Deseigne:2010gc}
Deseigne J, Dauchot O and Chat{\'e} H 2010 {\em Phys. Rev. Lett.\/} {\bf 105}

\bibitem{Palacci:2010hk}
Palacci J, Cottin-Bizonne C, Ybert C and Bocquet L 2010 {\em Phys. Rev.
  Lett.\/} {\bf 105} --

\bibitem{Theurkauff:2012ui}
Theurkauff I, Cottin-Bizonne C, Palacci J, Ybert C and Bocquet L 2012 {\em
  Phys. Rev. Lett.\/} {\bf 108} 268303

\bibitem{Deseigne2012}
Deseigne J, L{\'e}onard S, Dauchot O and Chat{\'e} H 2012 {\em Soft Matter\/}
  {\bf 8} 5629--5639

\bibitem{Bricard:2013jq}
Bricard A, Caussin J~B, Desreumaux N, Dauchot O and Bartolo D 2013 {\em
  Nature\/} {\bf 503} 95--98

\bibitem{Palacci:2013eu}
Palacci J, Sacanna S, Steinberg A~P, Pine D~J and Chaikin P 2013 {\em
  Science\/} {\bf 339} 936--940

\bibitem{Kumar:2014wr}
Kumar N, Soni H, Ramaswamy S and Sood A~K 2014 {\em Nat. Comm.\/} {\bf 5} 4688

\bibitem{peruani2006nonequilibrium}
Peruani F, Deutsch A and B{\"a}r M 2006 {\em Phys. Rev. E\/} {\bf 74} 30904

\bibitem{Grossman:2008is}
Grossman D, Aranson I~S and Ben-Jacob E 2008 {\em New Journal of Physics\/}
  {\bf 10} 023036

\bibitem{Henkes:2011ed}
Henkes S, Fily Y and Marchetti M~C 2011 {\em Phys. Rev. E\/} {\bf 84} --

\bibitem{Fily:2012tf}
Fily Y and Marchetti M~C 2012 {\em Phys. Rev. Lett.\/} {\bf 108}(23) 235702

\bibitem{Redner:2012vr}
Redner G~S, Hagan M~F and Baskaran A 2013 {\em Phys. Rev. Lett.\/} {\bf 110}
  055701

\bibitem{Weber:2013bj}
Weber C~A, Hanke T, Deseigne J, L{\'e}onard S, Dauchot O, Frey E and Chat{\'e}
  H 2013 {\em Phys. Rev. Lett.\/} {\bf 110} 208001

\bibitem{Solon:2013vr}
Solon A~P and Tailleur J 2013 {\em Phys. Rev. Lett.\/} {\bf 111}(7) 078101

\bibitem{Solon:2014tu}
Solon A~P, Chat\'e H and Tailleur J 2015 {\em Phys. Rev. Lett.\/} {\bf 114}(6)
  068101

\bibitem{Romenskyy2014}
Romensky M, Lobaskin V and Ihle T 2014 {\em Phys. Rev. E\/} {\bf 90}(6) 063315

\bibitem{Ballerini2007}
Ballerini M, Cabibbo N, Candelier R, Cavagna A, Cisbani E, Giardina I, Lecomte
  V, Orlandi A, Parisi G, Procaccini A, Viale M and Zdravkovic V 2008 {\em
  Proceedings of the National Academy of Sciences\/} {\bf 105} 1232--1237

\bibitem{Degond:2013wt}
Degond P, Appert-Rolland C, Pettr{\'e} J and Theraulaz G 2013 {\em Kinetic and
  Related Models\/} {\bf 6}(4) 809--839

\bibitem{Albi:2013ve}
Albi G, Balagu{\'e} D, Carrillo J~A and von Brecht J 2014 {\em SIAM J. Appl.
  Math.\/} {\bf 74} 794--818

\bibitem{Hanke2013}
Hanke T, Weber C~A and Frey E 2013 {\em Phys. Rev. E\/} {\bf 88}(5) 052309

\bibitem{aldana2007phase}
Aldana M, Dossetti V, Huepe C, Kenkre V and Larralde H 2007 {\em Phys. Rev.
  Lett.\/} {\bf 98} 95702

\bibitem{Kardar}
Kardar M 2007 {\em Statistical Physics of Particles\/} (Cambridge)

\bibitem{Watson1982}
Watson G 1982 {\em Journal of Applied Probability\/} {\bf 19} 265--280

\bibitem{Degond2013}
Degond P, Frouvelle A and Liu J~G 2013 {\em {J. Nonlinear Sci.}\/} {\bf 23}
  427--456

\bibitem{Chepizhko2014}
Chepizhko O and Kulinskii V 2014 {\em Physica A\/} {\bf 415} 493--502

\bibitem{Bird1970}
Bird G~A 1970 {\em Physics of Fluids (1958--1988)\/} {\bf 13} 2676--2681

\bibitem{Lobaskin2013}
Lobaskin V and Romenskyy M 2013 {\em Phys. Rev. E\/} {\bf 87}(5) 052135

\bibitem{Coburn2013}
Coburn L, Cerone L, Torney C, Couzin I~D and Neufeld Z 2013 {\em Physical
  Biology\/} {\bf 10} 046002

\end{thebibliography}
\providecommand{\newblock}{}

\end{document}